\documentclass[sigconf]{acmart}

\usepackage{algorithm,algpseudocode}
\usepackage{multirow}
\usepackage{tabularx}

\newdimen{\algindent}
\newcommand{\commentsymbol}{//}
\algrenewcommand\algorithmiccomment[1]{\hfill \commentsymbol{} #1}
\newdimen{\algindent}
\usepackage{varwidth}

\setcopyright{acmcopyright}
\copyrightyear{2022}
\acmYear{2022}
\setcopyright{rightsretained}
\acmConference[MOBILESoft '22]{IEEE/ACM 9th International Conference on Mobile Software Engineering and Systems}{May 17--24, 2022}{Pittsburgh, PA, USA}
\acmBooktitle{IEEE/ACM 9th International Conference on Mobile Software Engineering and Systems (MOBILESoft '22), May 17--24, 2022, Pittsburgh, PA, USA}
\acmDOI{10.1145/3524613.3527816}
\acmISBN{978-1-4503-9301-0/22/05}

\newcommand{\QD}{QuickDraw}
\newcommand{\dataName}{DoodleUINet}
\newcommand{\toolName}{PSDoodle}

\begin{document}

\title{\toolName{}: Fast App Screen Search via Partial Screen Doodle}

\author{Soumik Mohian}
\affiliation{%
  \department{Computer Science and Engineering Department}
  \institution{University of Texas at Arlington}
  \streetaddress{Box 19015}
  \city{Arlington}
  \state{Texas}
  \country{USA}
  \postcode{76019}
}
\email{soumik.mohian@mavs.uta.edu}

\author{Christoph Csallner}

\affiliation{%
  \department{Computer Science and Engineering Department}
  \institution{University of Texas at Arlington}
  \streetaddress{Box 19015}
  \city{Arlington}
  \state{Texas}
  \country{USA}
  \postcode{76019}
}
\email{csallner@uta.edu}

\begin{abstract}

Searching through existing repositories for a specific mobile app screen design is currently either slow or tedious. Such searches are either limited to basic keyword searches (Google Image Search) or require as input a complete query screen image (SWIRE). A promising alternative is interactive partial sketching, which is more structured than keyword search and faster than complete-screen queries. \toolName{} is the first system to allow interactive search of screens via interactive sketching. \toolName{} is built on top of a combination of the Rico repository of some 58k Android app screens, the Google \QD{} dataset of icon-level doodles, and \dataName{}, a curated corpus of some 10k app icon doodles collected from hundreds of individuals. In our evaluation with third-party software developers, \toolName{} provided similar top-10 screen retrieval accuracy as the state of the art from the SWIRE line of work, while cutting the average time required about in half.

\end{abstract}



\maketitle

\section{Introduction}

Searching through existing repositories for a specific mobile app screen design is currently either slow or tedious. Currently such searches are either limited to traditional keyword searches (e.g., via Google's image search) or require as input a complete query screen image (i.e., via the SWIRE line of work~\cite{huang2019swire,sain2020cross}) and are therefore slow and do not support well an interactive or iterative search style.

Having an effective and efficient search engine for mobile app screens can benefit many key software engineering tasks, including requirements gathering, understanding current market trends, analyzing features, providing inspiration to developers, and as a benchmark for evaluation~\cite{herring2009designpractise,eckert2000sourceinspiration}. Given the wide-spread and increasing use of mobile apps in a ``mobile first'' world and the resources spent on developing them~\cite{ines2017evalmobileinterface,hellmann2011ruletestUI}, such a screen search engine could have a large positive impact on many software developers and users. We are particularly focused on software developers with little to no UI/UX/design background. These users may only have a vague idea of the screen contents and are looking for inspiration from professional screen designs.

Several screen repositories exist, including websites such as Dribbble\footnote{\url{https://dribbble.com/}, accessed January 2022.} and Behance\footnote{\url{https://www.behance.net/}, accessed January 2022.}. Another repository is the Rico dataset curated from Android apps at runtime~\cite{deka2017rico}. Searching through this vast collection and finding desired example screens currently requires extensive effort via keyword-based search (e.g., for screen color, theme, date, and location) through several websites~\cite{ritchie2011d}. Moreover, novice users often fail to formulate good keyword queries and therefore do not get the intended search results~\cite{herring2009designpractise}.

Several researchers have proposed using visual, e.g., image- or sketch-based search methods because they are easy to use and fast to adopt~\cite{yeh2009sikuli}. For software development sketches are a natural fit, as sketches are a common form of visual representation, especially during early software development phases such as UI prototyping~\cite{Landay95Interactive, Newman99Sitemaps, Wong92Rough, Campos2007Practitioner,carter2010user}. 

\toolName{} is the first approach that supports interactive and iterative sketch-based screen search. \toolName{} uses a digital drawing interface with support for touchscreen devices of different resolutions and provides ease of use for the mouse. Using a digital drawing interface enables live search and user interaction. \toolName{} also does not suffer from the processing delays of paper-based approaches with their offline processing steps.

\begin{figure*}[h!t]
 \centering
 \includegraphics[width=\linewidth] {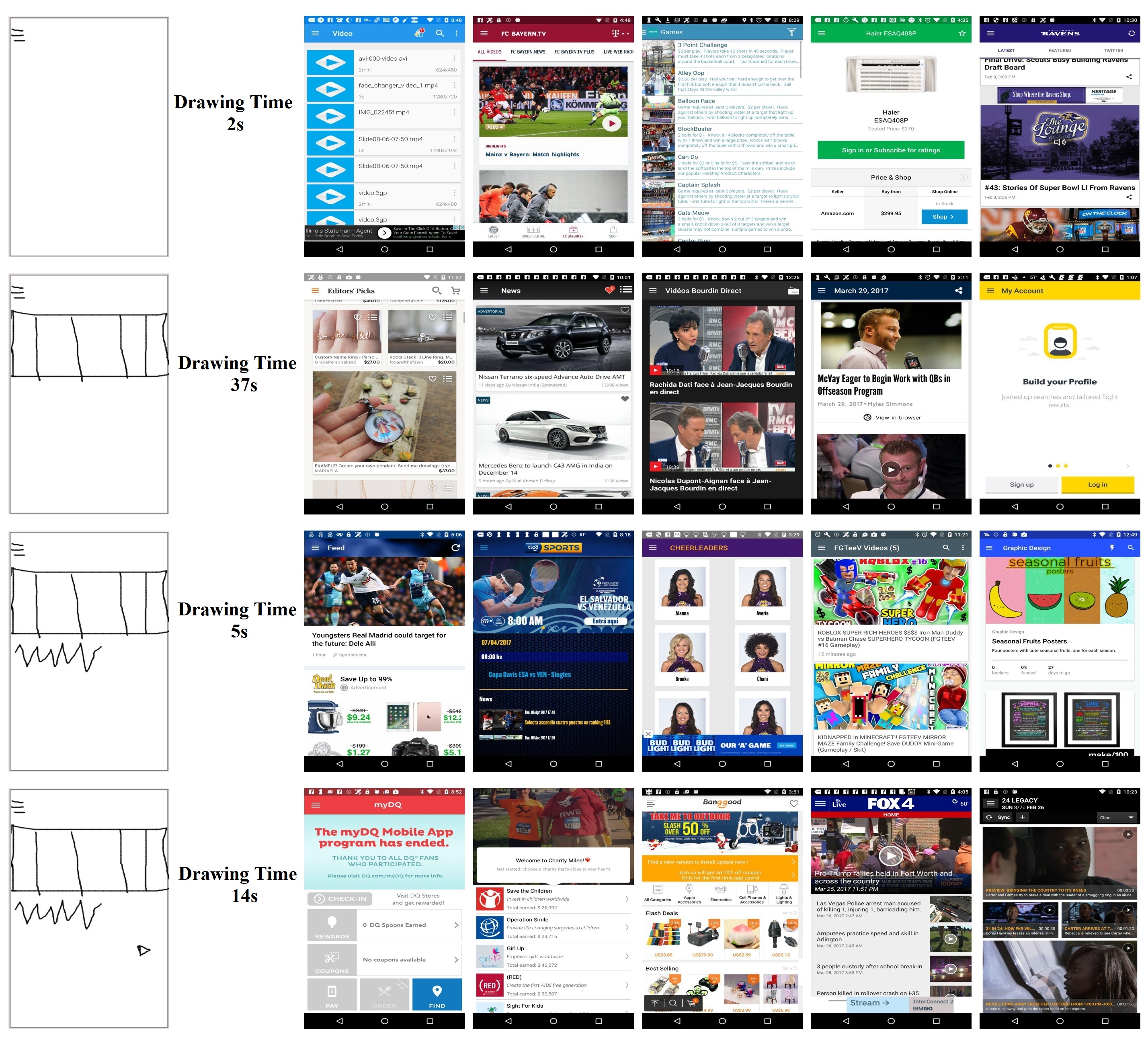}

 \caption{The second app screen search one of the study participants performed on \toolName{} (after finishing a 7-minute tutorial). Each row shows a user query sketch (1st~column) after adding one more UI~element, followed by \toolName{}'s top-5 (out of 58k) search results (in order). For each of these four queries, several of the result screens contain the sketched UI elements at about the sketched location. Drawing time contains all time from the user starting to work on the new UI element to the user indicating the new UI element sketch is finished. In each row \toolName{} returned the top-10 (ranked) result screens within 2~seconds (which includes a roundtrip from the user's machine to the AWS-hosted \toolName{}).}
 \label{fig:iterativeRetriveval}
\end{figure*}

Figure~\ref{fig:iterativeRetriveval} gives an overview of a sample \toolName{} search, starting in row 1 with the user sketching a ``hamburger''-style menu icon in the top left corner. Each of \toolName{}'s top-5 result screens contains a hamburger menu icon at about the sketched location. The second row shows the result of the user adding the doodle of a custom image below the hamburger icon, followed by two rows adding one more UI element doodle each.

\toolName{} employs deep learning to identify UI elements from drawing strokes.  \toolName{} fetches real-world UI examples from the Rico~\cite{deka2017rico} dataset based on UI element type, position, and element shape. It retrieves UI screens from the first UI element query element. \toolName{} updates the search result with the addition or removal of a UI element in the sketch.

At the same time \toolName{} provides search accuracy on par with state-of-the-art full-screen sketch approaches. We compared \toolName{} to state-of-the-art approaches by recruiting and observing 10 participants who used \toolName{} for the first time. We displayed a UI screenshot from Rico and instructed the participant to draw until the Rico screen appears in \toolName{}'s top search results. 88\% of the time \toolName{} retrieved and displayed the Rico target screen in its top-10 search results. A user usually spent an average of 107 seconds and drew an average of 5.5 elements during the process. This compared favourably with the most closely related tool SWIRE~\cite{huang2019swire}, which took 246 seconds to complete a sketch and took an average of 21.1 icon elements in each query drawing. To summarize, this paper makes the following major contributions.

\begin{itemize}

\item \toolName{} is the first tool that provides an interactive iterative search-by-sketch screen search. It is freely available online at: 
\url{http://pixeltoapp.com/PSDoodle/}

\item In our comparison with the state-of-the-art SWIRE line of work, \toolName{} achieved similar top-10 search accuracy while requiring less than 50\% of the time. 
\item All of \toolName{}'s source code, processing scripts, training data, and experimental results are available under permissive open-source licenses~\cite{soumikmohianuta_psdoodle_repo, soumik_mohian_DoodleUINet}. 
\end{itemize}

\section{Background}

Due to their wide use, we focus on Android apps and their common UI elements. Rico contains 72k unique app screens, collected from 9.3k Android apps from 27~app categories of the Google Play store~\cite{deka2017rico}. Rico ran (via ERICA~\cite{Erica}) each of these Android apps on modified Android classes to efficiently capture both screenshots and each screen's runtime UI view hierarchy. Rico thereby provides for each screenshot each UI element's Android class name, textual properties, x/y coordinates, and visibility.

A common challenge is understanding apps' custom UI elements (e.g., a clickable custom image used as an alternative implementation of a standard Android icon). To understand an UI element's intent beyond its Android class name, Liu et al. clustered 73k Rico screen elements by image similarity, the similarity of an element's surrounding text snippets, and similar code-based patterns ~\cite{liu2018learning}. This yielded 25 UI component types (e.g.: checkbox, icon, image, text, text button), 197 text button concepts (e.g.: no, login, ok), and 135 icon classes (e.g.: add, menu, share, star), with which Liu et al. labeled all screen elements in Rico.

\subsection{SWIRE: Offline Full-screen Search}

Most closely related to our work is SWIRE~\cite{huang2019swire}. SWIRE collected 3.8k low-fidelity full-screen Rico screen sketches from 4 experienced UI designers given a pre-defined drawing convention. Specifically, SWIRE instructs users to sketch each image as a crossed-out square (square borders plus diagonals) or as a square filled with a mountain outline. SWIRE users also represent any text with (a part of) the same 3-word template (Lorem ipsum dolor) or by squiggly lines. SWIRE trained a deep neural network on 1.7k Rico sketch-screenshot pairs created by 3~designers, yielding a top-10 screen retrieval accuracy of 61\% (i.e., in 61\% of cases the screenshot corresponding to the fourth designer's query sketch was one of SWIRE's top-10 search result screenshots).

SWIRE reflects a traditional paper-based design style. Users sketch with pen on paper inside an Aruco marker frame~\cite{garrido2014aruco} to streamline subsequent de-noising, camera angle correction, and projection correction. To change a sketch the user will likely have to start over. Even scanning or taking a snap once plus the subsequent processing steps requires significant time.

Recent SWIRE follow-up work reported a top-10 accuracy of 90.1\% \cite{sain2020cross}. While using different processing steps, at a high level it followed SWIRE's paper-based design style and thus faces similar challenges for interactive search.

\subsection{Google \QD{} \& \dataName{}}

Google's Quick, Draw! (``\QD{}'') offers some 50M doodles of 345 everyday categories, from ``aircraft carrier'' to ``zigzag''~\cite{ha2017neural,jongejan2016quick}. \QD{} doodles were sketched by anonymous website visitors, who were only given a one-word description of the thing to sketch. For each element category this yielded sketches performed in a wide variety of drawing styles. Given this diverse training set, \QD{} achieved solid doodle recognition accuracy for a wide range of sketching styles (e.g., earlier work reported some 70\% top-1 doodle recognition accuracy~\cite{quickdraw_rnn}). Internally \QD{} represents each doodle as a stroke sequence. Each stroke is a drawing from a start-touch to an end-touch event (e.g., mouse button press and un-press), represented by a sequence of straight lines.

\begin{figure}[h!t]
 \centering
 \includegraphics[width=\linewidth] {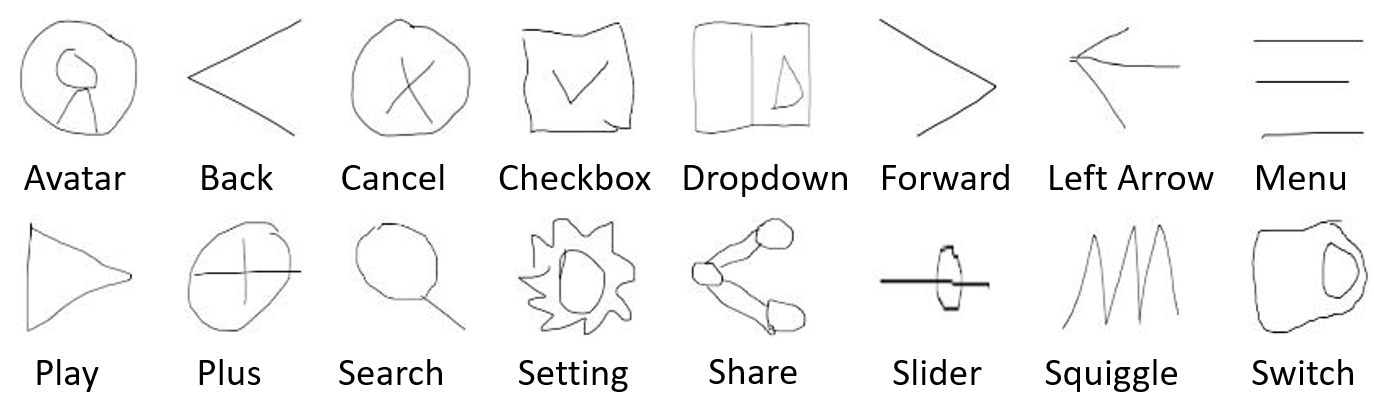}
 \caption{\toolName{}'s \dataName{} icons (1 sample per class).}
 \label{fig:crowd-icon-doodles-correct}
\end{figure}

\dataName{} offers some 11k crowdworker-created doodles of 16 common Android UI element categories~\cite{soumik_mohian_DoodleUINet}. Figure~\ref{fig:crowd-icon-doodles-correct} visualizes \dataName{}'s 16~UI element categories, spanning Android built-in element types (e.g., checkbox) and custom-designed images (e.g., avatar). \dataName{} doodles are stored in \QD{}'s format but do not overlap with \QD's doodle categories. In contrast to \QD{}'s flexible doodle recognition, the current version of \dataName{} focuses on a \emph{single drawing style} per element category (``stylized''), which it achieved by briefly presenting crowdworkers a stylized target image of the element they should sketch. (For example, in \dataName{} a UI element to reach ``settings'' currently always looks like a gear symbol.) Some 10k \dataName{} sketches are labeled ``correct'' (or similar-looking to the target image according to manual review) and some 1k are labeled ``incorrect''.

\section{Overview and Design}

\begin{figure}[h!t]
 \centering
 \includegraphics[trim=.2in .2in .2in .2in, clip, width=\columnwidth] {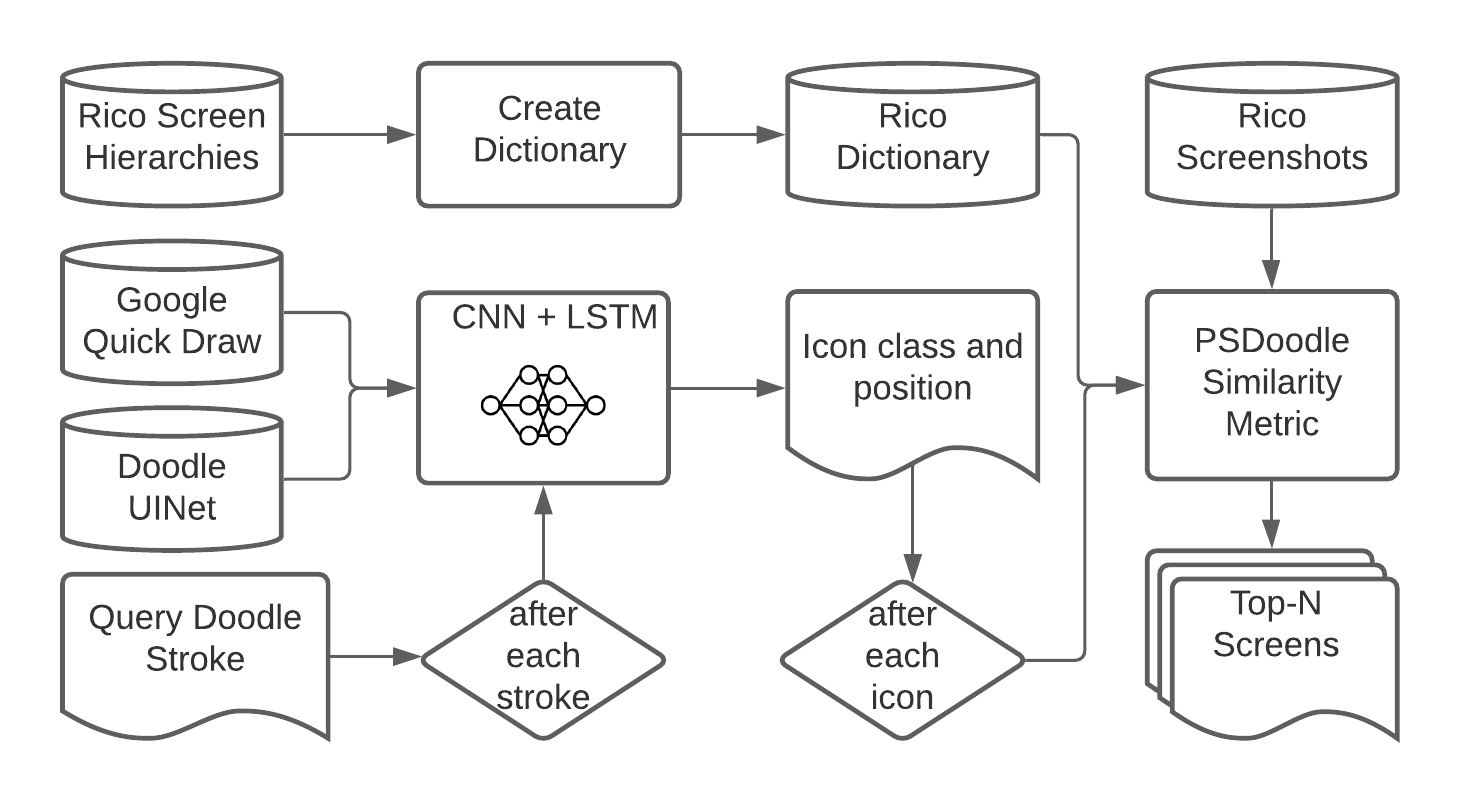}
 \caption{\toolName{} answers user queries via an offline-trained icon-level stroke-sequence recognizer and an offline-created hierarchy dictionary of 58k Rico screens.}
 \label{fig:architecture}
\end{figure}

Figure~\ref{fig:architecture} gives an overview of \toolName{}'s architecture. \toolName{} offers a drawing interface (bottom left) and recognizes a stroke sequence as an UI element via a deep neural network trained on \dataName{} and \QD{} doodles. After recognizing a new UI element, \toolName{} looks up the top-N matching screens in its dictionary of Rico screen hierarchies via \toolName{}'s similarity metric based on UI element shape, position, and occurrence frequency.

\subsection{Rico Screens \& UI Element Labels}

While the Rico paper mentions 72k screens, its dataset contains 66,261 screens. Given our UI element based search, \toolName{} cannot distinguish between screens with few UI elements (e.g., between two screens that only show a single large image). We thus exclude a Rico screen if the entire screen consists of a single text area (2,384 screens), a single image (561), single text plus single image (502), single webview (2,367), webview covering most of the screen area (1,433), or has no hierarchy information (888). (While a webview may contain an arbitrary webpage, Rico contains no information about this webpage.) This yields 58,126 Rico screens in \toolName{}.

\begin{table*}[htbp]

\begin{tabularx}{\textwidth}{p{6.5cm}Xl}
\hline
\textbf{Container's Android Class} & \textbf{Element's Android Class} & \textbf{New Label}  \\ \hline

CheckedTextView, AppCompatCheckedTextView, AppCompatCheckBox, CheckableImageView, AnimationCheckBox, CheckableImageButton, CheckBox, ColorableCheckBoxPreference & 
AppCompatCheckBox, PreferenceCheckbox, CheckboxChoice, CheckboxTextView, CenteredCheckBox, StyledCheckBox, CheckButton, AppCompatCheckBox, CheckBox, CheckBoxMaterial  & Checkbox    \\ 

\hline
 RangeSeekBar, SeekBar & 
 RangeSeekBar, TwoThumbSeekBar, EqSeekBar, PriceRangeSeekBar,  VideoSliceSeekBar, SliderButton  & Slider 
    \\ 
 \hline
 
 RatingBar & RatingWidget, RatingSliderView, RatingView, Rating & Star   \\  \hline
 
    SwitchCompat, Switch & 
  SwitchCompat, CustomThemeSwitchButton, BetterSwitch, LabeledSwitch,
  CustomToggleSwitch, CheckSwitchButton, CustomSwitch, MySwitch, SwitchButton, Switch & Switch    \\ \hline
  
n/a  &  CustomSearchView, SearchEditText, CustomSearchView, SearchBoxButton & 
 Search   \\ \hline
  
\end{tabularx}
\caption{47~patterns of fixes we have applied to ``input'' and ``image'' labels Liu et al. have applied to Rico UI elements on 3,317 screens. If Liu et al. labeled an element ``input'' or ``image'' and the element's type is a second column Android class or its direct container is a first column Android class, then we replace the label with the right column.}

\label{tab:additiona_riteria}
\end{table*}

For 3,317 Rico screens we noticed and fixed several inaccuracies in Liu et al.'s labeling of UI elements as ``input'' or ``image''. Table~\ref{tab:additiona_riteria} summarizes these fixes as 47~patterns. For example, we found that if Liu et al. labeled a UI element of Android class AppCompatCheckBox as an ``input'' then that UI element really looks like a ``checkbox''.

\subsection{Query Language: Stylized + Flexible Doodles}

While our long-term goal is to support every user and their preferred query styles (i.e., via an arbitrary mix of individual sketching styles, keywords, and structured query languages), \toolName{} focuses on sketch-only screen search. Specifically, \toolName{} combines the stylized \dataName{} sketch style with the flexible \QD{} sketch style. \QD{} has already validated that supporting both many categories and flexible drawing styles is possible using the \QD{} representation and classification \toolName{} has adapted. So migrating \toolName{} to support more UI element categories and a flexible drawing style is mostly a matter of collecting more training samples (and retraining \toolName{}'s neural net).

A key challenge not addressed by \QD{} is sketching deeply nested composite structures, which is common in app screens (e.g., a list of images plus text pairs in a container that is just one part of the screen). \toolName{} supports sketching such screens via its sequence-of-elements style. Specifically, the \toolName{} doodle classifier recognizes one UI element at a time. So once a user starts sketching a new doodle, \toolName{} treats each stroke as belonging to that UI element doodle, until the user indicates the UI element sketch is done. In that style it does not matter if the user first sketches a container or one of its (nested) UI elements, \toolName{} recognizes each separately and treats them as separate elements, allowing arbitrarily deeply nested container structures. 
\begin{figure}[h!t]
 \centering
 \includegraphics[width=\linewidth] {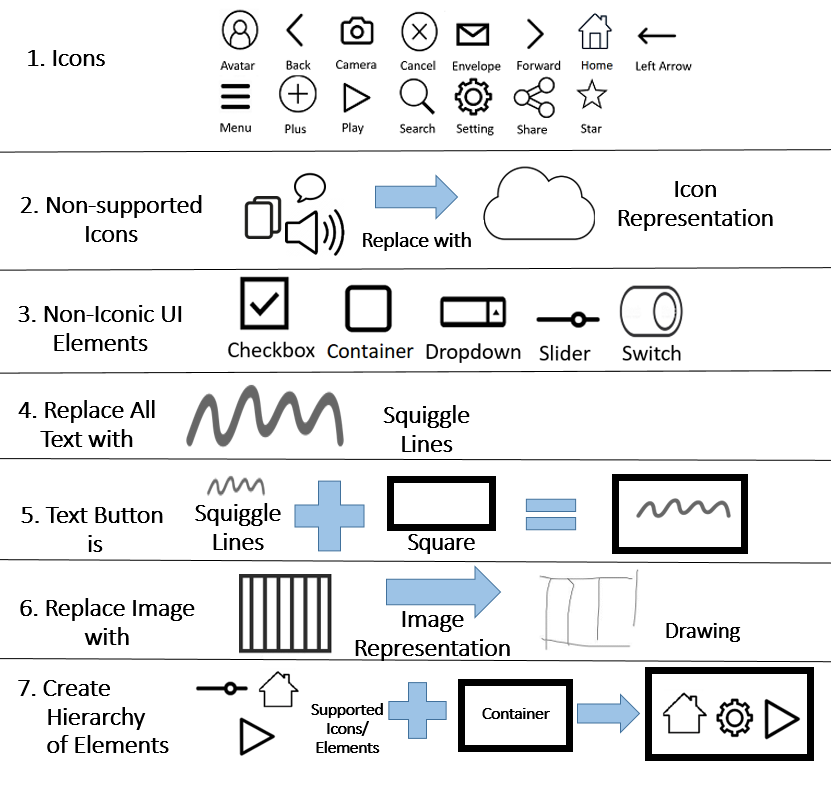}
 \caption{\toolName{}'s query language, as presented to users.}
 \label{fig:CheatSheet}
\end{figure}

\begin{figure}[h!t]
 \centering
 \includegraphics[width=.9\linewidth] {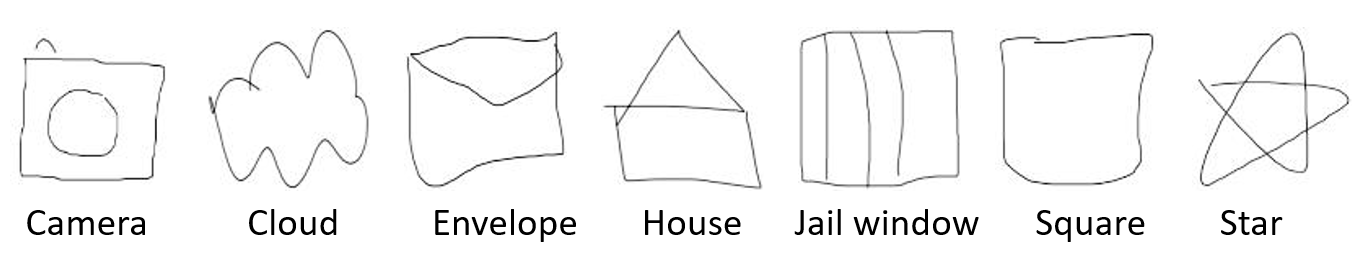}
 \caption{\toolName{}'s \QD{} icons (one sample per class).}
 \label{fig:qd-icon-doodles-correct}
\end{figure}

Figure~\ref{fig:CheatSheet} shows how \toolName{} presents its query language to its users (as a ``cheatsheet''). At the individual UI element level, \dataName{}~\cite{soumik_mohian_DoodleUINet} is a good fit for Android screen search as (according to the number of element labels inferred by Liu et al.~\cite{liu2018learning}) \dataName{} covers several of the most popular UI elements in Rico. Specifically, 11/16 of the stylized \dataName{} doodles look like the corresponding UI elements grouped and labeled by Liu et al. The other five either match SWIRE's language (squiggly line) or appear to be reasonable representations of common app concepts (dropdown, left arrow, slider, and switch). In addition to \dataName{}, we reviewed the \QD{} categories, looking for doodles that could be used to cover additional UI elements. We thereby identified 7~\QD{} classes (Figure~\ref{fig:qd-icon-doodles-correct} shows one sample each) that in our subjective judgement were a good match for UI sketching.

Besides the relatively close doodle-to-screen similarity of several classes (e.g., a sketched ``menu'' icon looks quite similar to an on-screen menu icon), \toolName{} follows SWIRE's approach of using a few placeholder elements to represent text and arbitrary images. For text \toolName{} uses a squiggly line (as SWIRE) and for an arbitrary image we use \QD{}'s ``jailwindow''. Furthermore, \toolName{} uses \QD{}'s cloud to represent a default (otherwise not directly-supported) icon and \QD{}'s square as a container.

Taken together, \toolName{} thereby covers the most common UI elements in Rico (in order) as follows (bold is from \dataName{}, italic from \QD{}): 
\textit{Container},                          
\textit{image}, 
\textit{icon} (a small interactive image),   
\textbf{text}, 
\textit{\textbf{text button}},                        
web view, 
input, 
list item, 
\textbf{switch} (a toggle element), 
map view, 
\textbf{slider}, and 
\textbf{checkbox}. 
Rico further sub-categorized the most popular icon types (\#3 in the above list) as 
\textbf{back}, followed by (in order) 
\textbf{menu}, 
\textbf{cancel} (close), 
\textbf{search}, 
\textbf{plus} (add), 
\textbf{avatar} (user head-shot type image), 
\textit{home}, 
\textbf{share}, 
\textbf{setting}, 
\textit{star}, 
edit, 
more, 
refresh, 
\textbf{forward}, and 
\textbf{play}.
\toolName{} further supports \textit{camera}, \textbf{dropdown}, \textit{envelope}, and \textbf{left arrow}.

Some popular UI elements can be treated as compound elements that can be composed of other more basic ones. \toolName{} supports one such case, i.e., the Android text button as ``text'' inside a ``square''. If a squiggle is inside a square and the square has no other nested UI elements then \toolName{} merges these two elements into a single (compound) element.

\subsection{UI Element Doodle Recognition}

\begin{figure}[h!t]
 \centering
 \includegraphics[width=.8\linewidth] {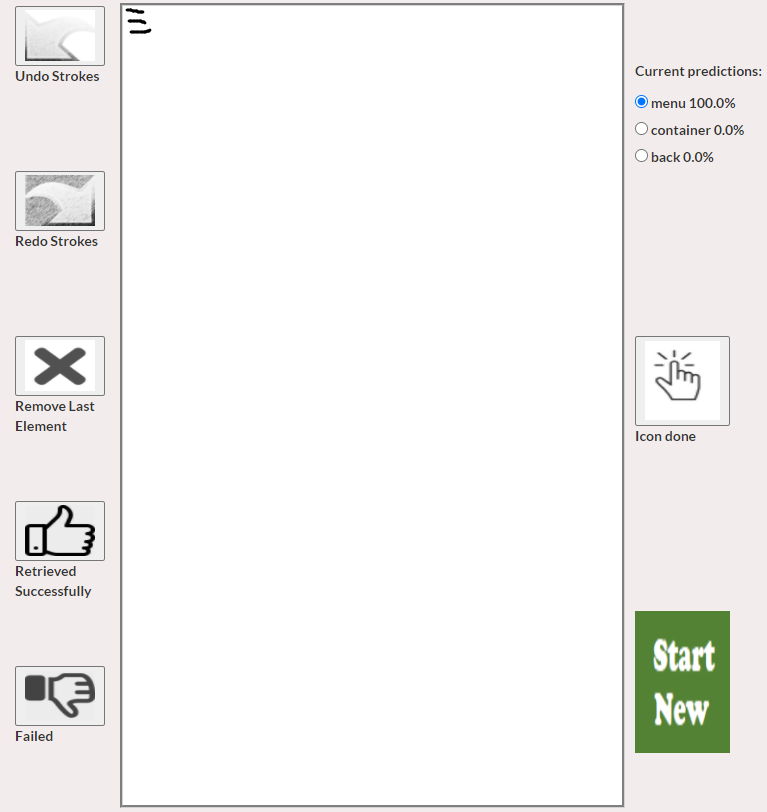}
 \caption{\toolName{} drawing UI, under which \toolName{} shows its current top-N Android search result screens (omitted).}
 \label{fig:searchInterface}
\end{figure}

Users draw on \toolName{}'s website via mouse or touch events. Users can undo or redo strokes and remove the last element doodle (Figure~\ref{fig:searchInterface} top left). Each time the user adds a stroke to the current doodle, \toolName{} shows its current top-3 UI element predictions (top right). A user can pick any of these three (and tap ``Icon done'') or continue editing the current UI element doodle. 

Once the user taps ``Icon done'', \toolName{} adds the sketched UI element to its search query, issues the query, and updates the display of its top-N Android search result screens.

To recognize a single UI element from strokes, we trained a deep neural network using \QD{}'s network architecture~\cite{quickdraw_rnn}, i.e., 
a 1-D convolutional neural network (CNN) layer (48 filters, kernel size 5) followed by
a 1-D CNN layer (kernel size 5, 64 filters),
a 1-D CNN layer (kernel size 3, 96 filters), 
3 Bi-LSTM layers, and 
a fully-connected layer. 

We used \dataName{} (some 600 doodles labeled ``correct'' for each of the 16 classes) plus a random 600-doodle sample of each of our 7 \QD{} classes.

We used transfer learning~\cite{torrey2010transfer} to pre-train the CNN layers for 23~\QD{} classes outside our 7~\QD{} classes. We then split our 23 classes into training and test samples (80\%/20\%) and trained the network for 24,893~steps, which yielded an accuracy of 94.5\% on the test data (which is similar to the 94.2\% accuracy a recent study achieved with 7-class \QD{} subset~\cite{Andersson2018SketchCW}).

To map an input stroke to \QD{}'s stroke-5 format~\cite{ha2017neural}, \toolName{} normalizes input stroke int locations to floats. Specifically, in the stroke-5 format~\cite{ha2017neural} each user input stroke is a sequence of points where each point is a tuple ($\Delta{} x, \Delta{}$ y, p1, p2, p3). Here p1 to p3 are binary sketch states after the current vertex (touching the canvas, raised from the canvas, done). $\Delta{} x$ and $\Delta{} y$ are integer pixel distances we normalize to floats (maintaining, among others, the number of vertices between input and normalized image).

\subsection{Searching Screens for UI Element Doodles}

After the user adds (or removes) a UI element, \toolName{} displays Rico screens that are similar to the current (partial) user screen sketch.
\toolName{} scores each of the 58k Rico screens based on how closely the screen matches the query doodles' presence, position, and shape. A key challenge is that a sketch is an abstract representation that is geometrically relatively far apart from its real-world counterpart. A UI element doodle thus will likely not be in the exact scale and position as it should appear on a UI screen. A similarity metric based on exact matching is thus likely to fail in sketch-based screen search.

\begin{figure}[h!t]
 \centering
 \includegraphics[width=\linewidth] {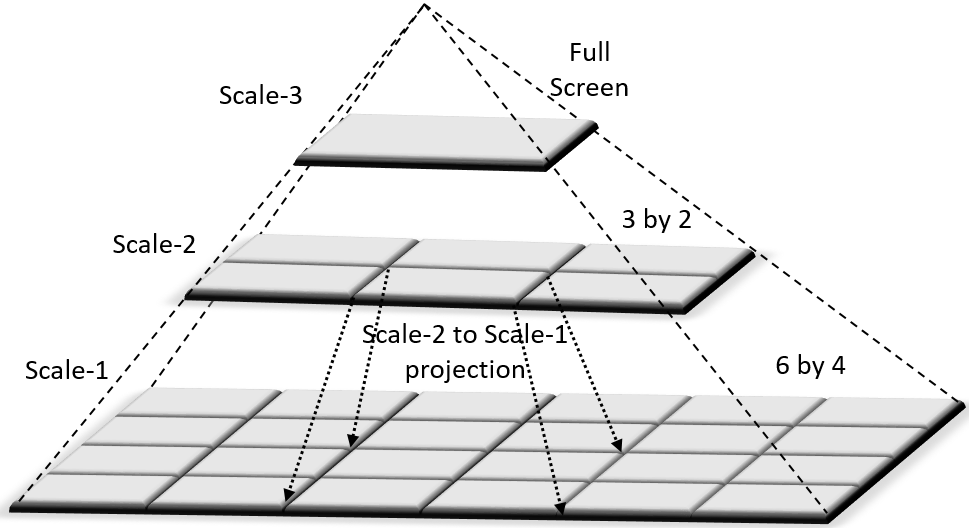}
 \caption{\toolName{}'s 3 levels of UI element search granularity.}
 \label{fig:AllScaleGrids}
\end{figure}

To address the issue, \toolName{} matches doodles (as recognized by the neural network after merging compound elements) with screen elements at different levels of screen resolution, starting at a fine-grained level but then backing up to more coarse-grained matches. Specifically, \toolName{}'s fine granularity scale-1 divides the canvas into 24~equal-sized rectangles (6 rows of 4 tiles each), scale-2 groups these into 6 rectangles (3 rows of 2 tiles), and finally scale-3 is a single rectangle. Figure~\ref{fig:AllScaleGrids} gives an overview of the different scale levels and how moving to a higher level widens the search area for a UI element match. 

Besides \toolName{}'s 3-level matching of a UI element's screen location, \toolName{} also matches the number of UI elements of a given class and takes into account how rare a UI element is within all screens. \toolName{} thus boosts the score of a rarer UI element as its presence may be more significant to the user. \toolName{} computes the inverse document frequency (IDF) of each UI element type in the 58k Rico screens (where a UI element type's IDF is larger if it appears on fewer screens).

\begin{algorithm}

\caption{scores screens by how closely they match the query doodles' size and location; $p_1$, $p_2$, $p_3$, $\Delta_w$, and $C_w$ are hyperparameters; $\Delta_A$ and $\Delta_C$ are UI element area- and count-differences $[0..1]$. }
  \label{alg:algorithm1}
\begin{algorithmic}[1] 
\State $res \leftarrow $ \{\} \Comment{Score per original screen}
\For{$doodles$ in $sketch$}  \Comment{Doodles of same type}
      \For{$screen$ in $dict[class(doodles)]$} \Comment{Screen + tiles}
      \State $z \leftarrow p_3$ \Comment{Screen's score} 
      \For{$tile$ in $tiles(screen)$} \Comment{Scale-1 tile} 
        \State $(A_o, C_o) \leftarrow elemAreaAndCount(screen[tile])$
        \If{$tile$ in $tiles(doodles)$} \Comment{Overlaps doodles}
        \State $(A_d, C_d) \leftarrow elemAreaAndCount(doodles[tile])$
        \State $\Delta_A \leftarrow 1 - |A_d-A_o|$  \Comment{Tile area: doodle vs orig}
        \State $\Delta_C \leftarrow max(0,1 - C_w \times |C_d-C_o|)$  \Comment{Counts}
        \State \begin{varwidth}[t]{\linewidth} $z \leftarrow z + 
        (p_1 \times { \dfrac{A_o}{C_o}}  \times {\dfrac{A_d}{C_d}}) + 
        (\Delta_w \!\times\! A_d \!\times\! \Delta_A \!\times\! \Delta_C$)
        \end{varwidth}
        \ElsIf{$tile$ in $neighbor(tiles(doodles))$}
            \State $z \leftarrow z + (p_2 \times { \dfrac{A_o}{C_o}})$
        \EndIf
        \EndFor
        \State $res[screen] \leftarrow res[screen] + (z \times idf[class(doodles)])$
      \EndFor
    \EndFor
 \State sort($res$)   \Comment{Retrieved screens by score} 
\end{algorithmic}
\end{algorithm}

For fast screen retrieval, \toolName{} maintains a dictionary of Rico's 58k (``original'') screens. This dictionary maps each of \toolName{}'s 24 UI element types to a list of screens, where each screen lists for each of its tiles the percentage of the tile's area ($A_o$) being covered by how many ($C_o$) instances of that UI element type. Algorithm~\ref{alg:algorithm1} summarizes \toolName{}'s screen similarity scoring as pseudo code. The algorithm iterates through the query sketch's UI element doodles, one element class at a time (e.g., starting with all of the query sketch's squiggly line doodles taken as a group). For each element doodle group, the algorithm looks up the Rico screens that contain at least one instance of the doodle group's element type (Line~3). For each matching Rico screen, we then iterate over the screen's scale-1 tiles that contain the given doodle type (Line~5) .

For each such original Rico screen tile that contains the doodled UI element type there are now three cases. First, if the original screen tile matches a tile of the doodle group then we have a scale-1 match and compute the percentage of that tile's area ($A_d$) being covered by how many ($C_d$) doodles of that UI element type (Line 8). Then we compute the difference in the tile's area coverage percentages between doodles and original screen elements (Line~9) and the difference in how many doodles vs how many original screen elements are in that tile (Line~10). We add the resulting tile score to its screen's overall score (Line~11).

In the second case (Line~12), the screen tile containing a UI element of the doodle group's class does not overlap with any tile of the doodle group. In this case \toolName{} backs up to its scale-2 search and checks if this screen tile overlaps with any direct neighbor tiles of the doodle group's tiles. If this is the case the screen gets a smaller score boost. Finally, if there is neither a scale-1 nor a scale-2 match, then the screen's score remains unchanged.

\subsection{Hyperparameter Optimization}

Algorithm~\ref{alg:algorithm1} mentions five hyperparameters, three for the scale levels ($p_1$, $p_2$, $p_3$) plus $\Delta_w$ for the weight difference and $C_w$ for the occurrence difference. To find optimal values for these hyper-parameters we collected 30 sketches from 5 computer science graduate students.

The collected sketches represent 30 different Rico screens that have at least two \toolName{}-supported icons. None of these screenshots or sketches were used for the tool evaluation. An exhaustive search with GridSearchCV of scikit-learn~\cite{sklearn_api} to get a high score and top-rank for the target screen yielded the optimized values
$p_1=39$,
$p_2=8$,
$p_3=9$, 
$\Delta_w=0.4$, and
$C_w=11$.

A closer look at the hyper-parameters indicates that they give more weights to scale-1 matches compared to the other two scales. With 24 girds, a scale-1 match implies higher location similarity, which we intuitively expect to yield a higher screen score.

\begin{table*}[htbp]
\begin{center}
  \caption{Strokes per doodle in \toolName{}'s data sets (left) and its partial doodle recognition trained on 80\% classifying the other 20\% (the test doodles) (right): 1st stroke at which \toolName{} ranks a doodle's correct class first (top-1) and within the top-3; W~=~test doodles \toolName{} classifies wrongly at the last/all strokes; W*~=~after retraining from scratch adding samples that remove the outer boundary of avatar, cancel, checkbox, plus;
  m~=~median;
  l~=~min;
  h~=~max;
  SD~=~standard deviation;
  cnt~=~count.}
  
  \label{tab:strokeStats}
  \begin{tabular}{lrrrrr|r|rrrrrrrrr|rrrrrrrrr}
   \hline
    \multicolumn{1}{l} {\textbf{Category}} &
    \multicolumn{5}{c|} {\textbf{Strokes in 100\%}} &
    \multicolumn{1}{l|} {\textbf{20\%}} &
    \multicolumn{5}{c} {\textbf{1st stroke to top-1}} &
    \multicolumn{2}{c} {\textbf{W} [\%]} &
    \multicolumn{2}{c|} {\textbf{W*} [\%]} &
    \multicolumn{5}{c} {\textbf{1st stroke to top-3}}  &
    \multicolumn{2}{c} {\textbf{W} [\%]} &
    \multicolumn{2}{c} {\textbf{W*} [\%]} 
    \\
    & avg & m & l & h & SD & cnt & avg & m & l & h & SD & lst & all & lst & all & avg & m & l & h & SD & lst & all & lst & all\\
      \hline
      
Camera & 3.5 & 3 & 1 & 9 & 1.1 & 143 & 2.3 & 2 & 1 & 5 & 0.7 & 6 & 4 & 13 & 8 & 1.8 & 2 & 1 & 5 & 0.8 & 1 & 1 & 1 & 1 \\
Cloud & 1.4 & 1 & 1 & 24 & 1.4 & 154 & 1.1 & 1 & 1 & 3 & 0.4 & 12 & 10 & 4 & 3 & 1.1 & 1 & 1 & 3 & 0.4 & 3 & 2 & 1 & 1 \\
Envelope & 2.1 & 2 & 1 & 9 & 1.1 & 145 & 1.9 & 2 & 1 & 9 & 1.1 & 7 & 6 & 6 & 6 & 1.3 & 1 & 1 & 4 & 0.6 & 0 & 0 & 0 & 0 \\
House & 3.5 & 3 & 1 & 23 & 2.2 & 135 & 2.1 & 2 & 1 & 5 & 0.9 & 11 & 9 & 1 & 1 & 2.0 & 2 & 1 & 10 & 1.1 & 3 & 2 & 1 & 1 \\
Jail-win & 5.8 & 5 & 2 & 20 & 1.7 & 143 & 3.4 & 3 & 2 & 8 & 1.1 & 7 & 6 & 1 & 1 & 2.9 & 3 & 1 & 8 & 1.2 & 2 & 1 & 1 & 0 \\
Square & 1.3 & 1 & 1 & 4 & 0.7 & 147 & 1.1 & 1 & 1 & 3 & 0.3 & 2 & 2 & 2 & 1 & 1.1 & 1 & 1 & 3 & 0.3 & 1 & 1 & 1 & 1 \\
Star & 1.4 & 1 & 1 & 10 & 1.0 & 148 & 1.1 & 1 & 1 & 4 & 0.5 & 1 & 1 & 1 & 1 & 1.1 & 1 & 1 & 2 & 0.3 & 0 & 0 & 0 & 0 \\

      \hline

Avatar & 3.8 & 4 & 1 & 8 & 0.8 & 136 & 2.9 & 3 & 1 & 7 & 0.9 & 9 & 8 & 2 & 2 & 2.3 & 2 & 1 & 6 & 0.7 & 1 & 1 & 0 & 0 \\
Back & 1.1 & 1 & 1 & 12 & 0.7 & 122 & 1.0 & 1 & 1 & 2 & 0.2 & 3 & 2 & 5 & 3 & 1.0 & 1 & 1 & 2 & 0.1 & 0 & 0 & 4 & 3 \\
Cancel & 3.1 & 3 & 2 & 12 & 0.5 & 127 & 2.6 & 3 & 1 & 5 & 0.6 & 25 & 21 & 3 & 2 & 2.2 & 2 & 1 & 4 & 0.6 & 2 & 1 & 1 & 1 \\
Checkbox & 2.8 & 2 & 1 & 51 & 2.6 & 134 & 2.3 & 2 & 1 & 10 & 1.4 & 13 & 13 & 4 & 1 & 1.7 & 1 & 1 & 7 & 1.1 & 3 & 2 & 1 & 1 \\
Drop-dwn & 4.5 & 3 & 2 & 43 & 3.3 & 133 & 2.8 & 2 & 2 & 6 & 1.0 & 3 & 2 & 5 & 3 & 1.9 & 2 & 1 & 5 & 0.7 & 2 & 2 & 2 & 2 \\
Forward & 1.1 & 1 & 1 & 17 & 0.8 & 122 & 1.0 & 1 & 1 & 1 & 0.0 & 0 & 0 & 2 & 2 & 1.0 & 1 & 1 & 1 & 0.0 & 0 & 0 & 1 & 0 \\
Left arrow & 2.2 & 2 & 1 & 10 & 0.8 & 123 & 2.0 & 2 & 1 & 4 & 0.5 & 13 & 12 & 3 & 2 & 1.9 & 2 & 1 & 4 & 0.6 & 5 & 5 & 1 & 1 \\
Menu & 3.2 & 3 & 2 & 16 & 1.1 & 126 & 2.0 & 2 & 1 & 3 & 0.4 & 0 & 0 & 3 & 2 & 1.8 & 2 & 1 & 3 & 0.4 & 0 & 0 & 0 & 0 \\
Play & 1.7 & 1 & 1 & 15 & 1.2 & 127 & 1.4 & 1 & 1 & 2 & 0.5 & 5 & 4 & 9 & 3 & 1.3 & 1 & 1 & 3 & 0.5 & 2 & 1 & 3 & 1 \\
Plus & 3.1 & 3 & 2 & 11 & 0.7 & 120 & 2.2 & 2 & 1 & 6 & 0.9 & 8 & 8 & 12 & 8 & 1.8 & 2 & 1 & 4 & 0.8 & 2 & 2 & 3 & 3 \\
Search & 2.3 & 2 & 1 & 13 & 1.0 & 122 & 2.0 & 2 & 1 & 3 & 0.3 & 2 & 2 & 9 & 7 & 1.9 & 2 & 1 & 3 & 0.4 & 1 & 1 & 3 & 2 \\
Setting & 5.6 & 2 & 1 & 61 & 7.1 & 111 & 3.1 & 2 & 1 & 34 & 4.0 & 11 & 7 & 5 & 5 & 2.5 & 2 & 1 & 24 & 2.4 & 1 & 0 & 1 & 1 \\
Share & 7.0 & 7 & 1 & 23 & 1.1 & 117 & 3.7 & 3 & 2 & 13 & 1.3 & 3 & 1 & 5 & 3 & 2.8 & 3 & 2 & 7 & 0.8 & 1 & 1 & 1 & 0 \\
Slider & 2.6 & 3 & 1 & 19 & 1.0 & 134 & 1.9 & 2 & 1 & 4 & 0.8 & 4 & 4 & 2 & 2 & 1.6 & 2 & 1 & 4 & 0.6 & 1 & 1 & 1 & 1 \\
Squiggle & 1.3 & 1 & 1 & 52 & 2.5 & 144 & 1.1 & 1 & 1 & 4 & 0.4 & 3 & 1 & 0 & 0 & 1.0 & 1 & 1 & 4 & 0.3 & 1 & 1 & 0 & 0 \\
Switch & 3.1 & 2 & 1 & 16 & 1.6 & 137 & 2.3 & 2 & 1 & 6 & 1.0 & 6 & 4 & 6 & 5 & 1.6 & 1 & 1 & 5 & 0.8 & 0 & 0 & 1 & 1 \\

    \hline
\end{tabular}
\end{center}
\end{table*}

\section{Evaluation}

We evaluated \toolName{}'s recognition accuracy of partial UI element doodles, its top-10 retrieval accuracy of partial screen sketches, and its screen retrieval time using the following research questions.

\begin{description}
  \item[RQ1] Can \toolName{} recognize partial UI element sketches?
  \item[RQ2] Can \toolName{} achieve similar top-10 accuracy as state-of-the-art screen search approaches?
  \item[RQ3] At a similar accuracy level, how many 
 
  UI elements did participants sketch in \toolName{} compared to state-of-the-art complete-screen sketch approaches?

  \item[RQ4] At a similar accuracy level, can \toolName{} retrieve screens faster than state-of-the-art approaches?
\end{description} 

Following the most closely related work~\cite{huang2019swire,sain2020cross}, we evaluated screen search performance by measuring top-k (screen) retrieval accuracy. We thus showed a participant a target screen to sketch and measured where in the result ranking the target screen appears. Top-k retrieval accuracy is the most common metric for sketch-based image retrieval tasks and correlates with user satisfaction~\cite{huffman2007well}.

Specifically, we evaluated screen search (in RQ2, RQ3, and RQ4) with 30 ``target'' Rico screens, which we selected as follows. To ensure the target Rico screens contain at least some UI elements \toolName{} supports, we removed from Rico's 58k screens those that contain less than two \toolName{}-supported UI elements, yielding 50,113 screens. From these 50k we randomly picked 30 screens, of which 26 were also in the SWIRE dataset.

We recruited 10 Computer Science students (all ages under 30). None of the participants had any formal UI/UX design training. All participants had heard about mobile app development principles before the study. For diversity, we recruited 5 participants (1~female, 4~male) without plus 5 (2~female, 3~male) with some prior mobile app development experience. Each student was compensated with USD~10 and used \toolName{} for the first time.

For the experiment, we used \toolName{}'s regular setup as a website hosted on an Amazon AWS EC2 general purpose instance (t2.large) with two virtual CPUs and 8 GB of RAM. Each participant interacted with \toolName{} over the internet from their personal machine (i.e., a laptop or desktop computer). Each participant first spent an average of 9 minutes on the \toolName{}'s interactive tutorial, which covers \toolName{}'s visual language, how and where to draw, how to access the cheat sheet (Figure~\ref{fig:CheatSheet}), how to see the search results, and when to stop the search (\url{http://pixeltoapp.com/toolIns/}). 

After the tutorial each user was instructed to sketch at least 3 screens. The \toolName{} website recorded their drawings, drawing time, and query results. Throughout the experiments, we observed participants' performance via screen sharing but did not otherwise interact with them (e.g., to coach them on how to use \toolName{}). All records are available in the \toolName{} repository.

\subsection{RQ1: Recognizing Partial Icon Doodles}

Table~\ref{tab:strokeStats} compares the number of per-doodle strokes in \toolName{}'s data sets with the number of strokes \toolName{} takes to correctly classify a doodle. For the latter Table~\ref{tab:strokeStats} lists two criteria, ranking the correct class first (middle columns) and ranking the correct class in the top-3 (right columns). In the experiments participants often kept sketching until \toolName{} ranked the correct class top-1 but sometimes stopped sketching after selecting the correct class from the top-3 predictions the \toolName{} UI provides after each stroke.

For both criteria (top-1 and top-3), users can often transmit their intent to \toolName{} with fewer than a doodle's full set of strokes. For example, while the average avatar doodle contains 3.8 strokes, \toolName{} ranks avatar top-1 on average after 2.9 strokes and top-3 after just 2.3 strokes. Several other classes have similarly large reductions in average stroke counts.

\begin{figure}[h!t] 
 \centering
 \includegraphics[width=\linewidth] {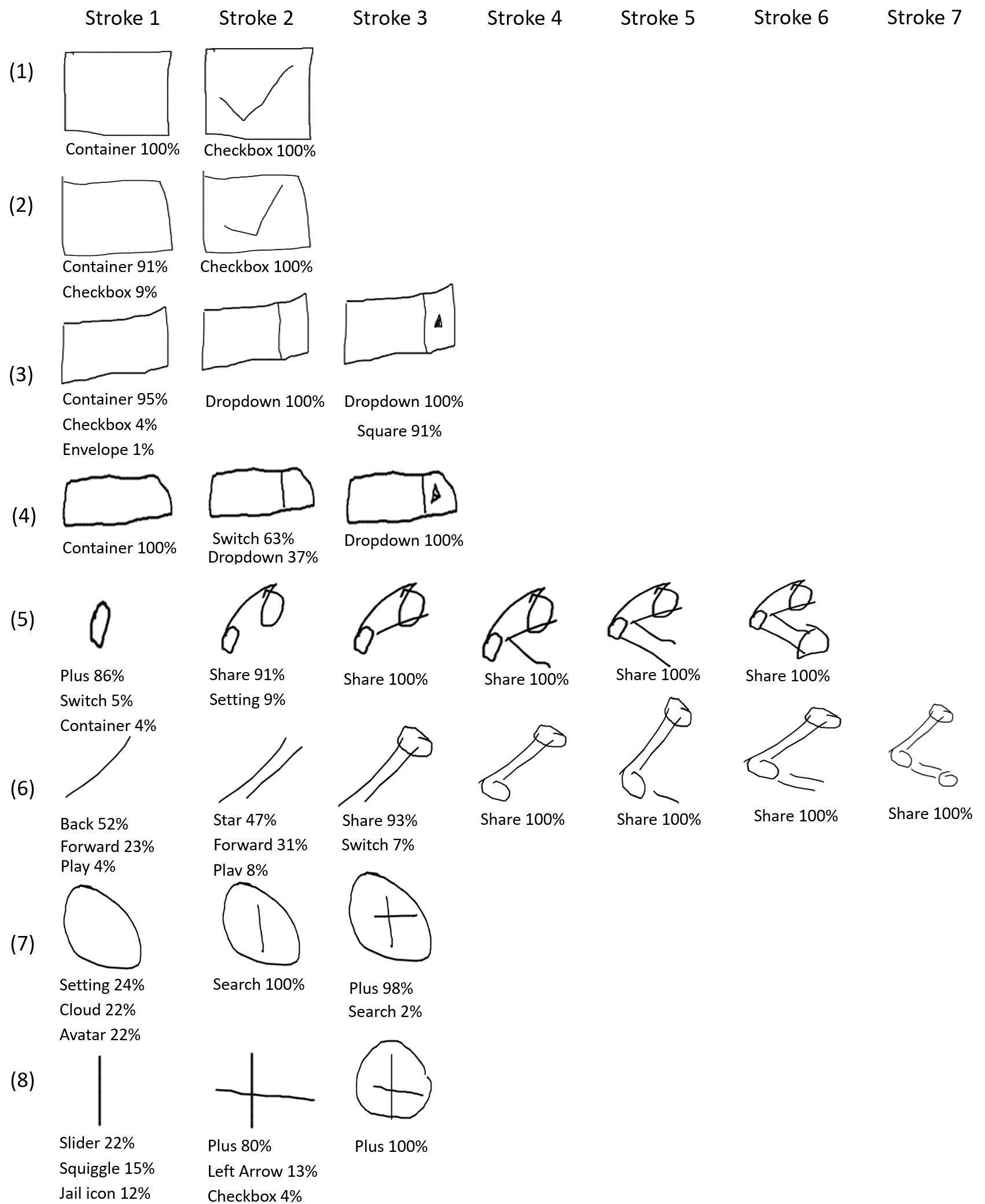}
 \caption{Two random samples (from the 20\% of doodles) from 4 categories. Below each incremental stroke is \toolName{}'s confidence for its current top predictions. For 6 of these 8 doodles \toolName{} reached the correct prediction before the last stroke, allowing the user to communicate their intent without finishing the doodle.}
 \label{fig:icon-predict-by-stroke}
\end{figure}

To visualize \toolName{} performance on concrete examples, Figure~\ref{fig:icon-predict-by-stroke} displays \toolName{}'s current prediction after each stroke of 8~randomly sampled drawings from 4~categories. For example, in the fifth row \toolName{} ranks the doodle's correct class top-1 after only two strokes of the doodle's six total strokes. Such an early correct classification allows the user to quickly move on to the next doodle, thereby saving time and receiving query results faster.

\begin{figure}[h!t]
 \centering
 \includegraphics[width=\linewidth]{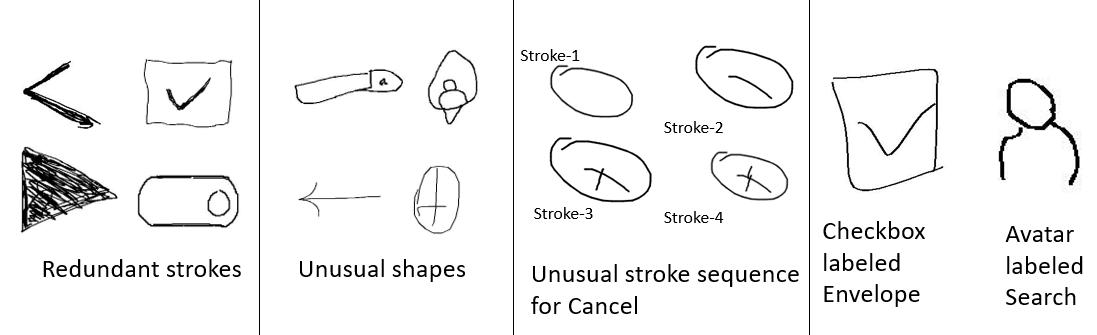}
 \caption{Test set drawings the network misclassified.}

 \label{fig:icon-predict-fails}
\end{figure}

We also inspected a random sample of 35 of the 192~test set sketches \toolName{} misclassified after the last stroke. Among these 35 samples we found four patterns, i.e., being an outlier due to using a large number of strokes (compared to the doodle class's average) for 7/35 samples, using unusual strokes (such as a squiggle during drawing using more vertices compared to the class's vertex average) or stroke sequence for 10/35 samples, deviation from the class's common shape (14/35), and resembling another category (4/35). Figure~\ref{fig:icon-predict-fails} shows examples of these four categories.

\subsection{RQ2: Top-10 Screen Search Accuracy}

For this experiment participants were instructed to sketch with the goal of using \toolName{} to retrieve a given Rico ``target'' screen. We then measured how quickly this target Rico screen appeared in \toolName{}'s top-10 search results. We asked participants to use the tool at least 3 times, with 4 users attempting one additional sketch each, yielding 34 screen sketches of 30 Rico screens. For these 34 sketches, 30 times the target UI screen appears in the top-10 results, yielding a top-10 accuracy of 88.2\%. (Since \toolName{} shows rows of result screens similar to Google's image search, top-1 accuracy is less relevant.) \toolName{}'s top-10 accuracy is significantly higher than SWIRE's and remains similar to a recent SWIRE follow-up work by Sain et. al~\cite{sain2020cross}, which reported 90.1\% top-10 accuracy for SWIRE sketches.

\begin{figure}[h!t]
 \centering
 \includegraphics[width=.8\linewidth] {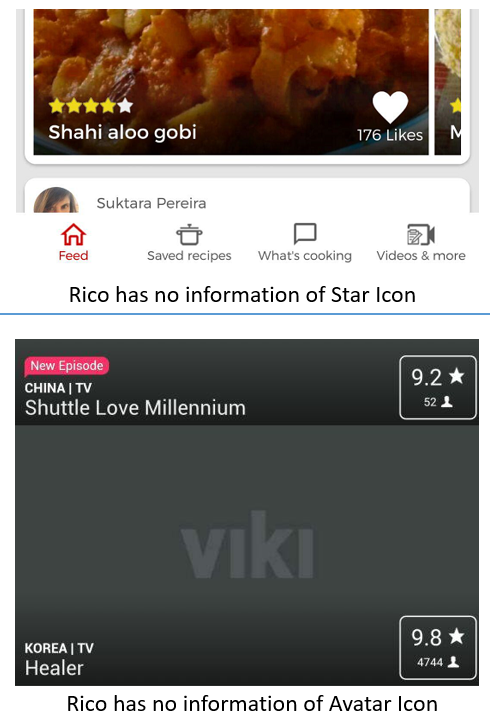}
 \caption{\toolName{} fails to rank these two target screens (excerpted) in its top-10 search results, due to their Rico hierarchy having no information about an icon drawn by a user.}
 \label{fig:Search_By_Skecth_Fails}
\end{figure}

We manually checked each case where \toolName{} failed to rank the target screen in the top-10. Figure~\ref{fig:Search_By_Skecth_Fails} shows excerpts of two such screens. In both the Rico hierarchy does not contain the correct label of a user-drawn icon. Such cases could be reduced by further improving Rico's UI element clustering and classification. Another case stems from human error (i.e., participant 9 in Table~\ref{tab:participantsScore}) because the user selected the wrong doodle category from \toolName{}'s top-3 prediction. Such human errors may become less common once users become more experienced with using \toolName{}.

\subsection{RQ3: Search With Partial-screen Sketches}

In addition to recognizing partial UI element sketches, \toolName{} also supports an iterative search style where a user refines the search results one UI element at a time. As the SWIRE-style approaches process complete-screen sketches, this research question quantifies the difference in UI elements a user has to draw to perform a successful screen search. Answering this question is made easier by the earlier experiment yielding a similar top-10 accuracy for \toolName{} and the most-accurate SWIRE-style approach.

In the 30 target Rico screens participants used, the average UI element count was 21.1 with a median of 19 (low 14, high 35, and standard deviation 5.4). SWIRE instructs users to sketch all screen elements, so a SWIRE sketch has a similar number of UI elements. In the participants' 34~\toolName{} sketches of these 30 screens the average UI element count was significantly lower at 5.5 with a median of 5 (low 3, high 9, and standard deviation 1.8).

We also tested how the most-accurate SWIRE-style approach would perform on the partial screen sketches our participants produced with \toolName{}, by training their network to the reported 90.1\% top-10 and 67\% top-1 accuracy for SWIRE sketches~\cite{sain2020cross}. We thus converted our 34 participant sketches from \QD{}'s sequence-of-stroke format to SWIRE-style black/white bitmaps (and included them in the \toolName{} repository). We removed the 8 participant sketches that used jail-window, as SWIRE uses a different placeholder for images. For all resulting 26 query sketches, the SWIRE follow-up failed to fetch the target screen in the top-10.

\subsection{RQ4: Interactive and Fast Screen Retrieval}

In our experiments we told participants to search via sketching for 3 minutes and stop sketching if the target screen appears in \toolName{}'s top-10 result. We recorded how long each search session took. SWIRE reports that the sketching alone of each SWIRE sketch of a Rico screen took an average of 246 seconds.

\begin{table}[htbp]
\begin{center}
  \caption{Time (seconds) a participant (P) took to iteratively sketch a target screen and retrieve result screens. The final target screen ranking (r) was top-10 accurate in 88\% of cases.}

  \label{tab:participantsScore}
  \begin{tabular}{r|rr|rr|rr|rr}
   \hline
    \multicolumn{1}{l|} {\textbf{P}} &
    \multicolumn{2}{c|} {\textbf{Target 1}} &
    \multicolumn{2}{c|} {\textbf{Target 2}} &
    \multicolumn{2}{c|} {\textbf{Target 3}} &
    \multicolumn{2}{c} {\textbf{Target 4}} 
    \\
    & t & r & t & r & t & r & t & r  \\
      \hline
1 & 55 & 2 & 51 & 1 & 134 & 8 & - & - \\
2 & 86 & 3 & 259 & 7 & 206 & 3 & - & - \\
3 & 134 & 3 & 97 & 4 & 63 & 2 & - & - \\
4 & 85 & 5 & 75 & 1 & 64 & 5 & - & - \\
5 & 202 & 3 & 60 & 1 & 105 & 14491 & - & - \\
6 & 127 & 2 & 119 & 1 & 109 & 9 & - & - \\
7 & 44 & 2 & 98 & 1 & 38 & 10 & 39 & 1 \\
8 & 168 & 10 & 248 & 1 & 103 & 1 & 73 & 5 \\
9 & 158 & 31 & 46 & 61 & 97 & 31 & 40 & 3 \\
10 & 30 & 8 & 138 & 1 & 147 & 1 & 158 & 6 \\
    \hline
\end{tabular}
\end{center}
\end{table}

Table~\ref{tab:participantsScore} lists the total time of the search and sketch session for each of the experiment's 34 sessions, together with the final rank of the target Rico screen in \toolName{}'s search results. Total sketch and search times per search session varied from 30 to 259 seconds. While achieving similar top-10 accuracy, most of these session times were significantly shorter than the 246 second average of SWIRE for sketching only. Most of \toolName{}'s session times were also significantly shorter than the 180 seconds target provided to participants.

\toolName{} is deployed in AWS and supports interactive search. In our experiments there was less than 2s delay between the user submitting a search query (e.g., by pressing “icon done”) to the update of the top-10 result screens on the user’s \toolName{} website. Besides communication to and from AWS, the main time components were sketch recognition (below 0.1s) as well as screen similarity calculation and screen ranking (below 1s).

\subsection{High-level Feedback from Participants}

Two of our 10 participants opted out of our post-evaluation survey, leaving us with 8 completed surveys. In one question we asked how participants prefer to sketch. 
Three preferred touch to sketch on a larger device such as an iPad, 
two preferred sketching on paper and taking a picture with their phone, 
two wanted to use a mouse to sketch on a non-touch device,
one wanted to touch to sketch on a smaller device such as a smartphone, and 
none wanted to scan a paper-based sketch. Overall participants preferred device-based over paper-based sketching by 3:1.

In another question we asked participants to choose from three sketch-based search tool options. Two participants voted for an approach that shows its search result only after finishing a complete screen (containing all the UI elements that the screen should have in the app). The other six participants preferred a search tool that shows live search results (i.e., search results that update when adding or removing a UI element). None of the participants picked the third option, a tool that only shows its results after sketching a partial screen containing several icons.

\subsection{Relaxing \toolName{}'s Query Language}

In informal but more concrete feedback, participants explained how they sometimes struggled with the four \toolName{}'s icon classes whose shapes include outer boundaries such as avatar's ``outer ring''. These classes were avatar, cancel, checkbox, and plus. 

While some participants preferred to sketch such icons without these outer boundaries, \toolName{}'s training data sets contained only few such samples. 

To address this issue, after the experiments with participants we created additional samples from \toolName{}'s existing samples, by identifying and removing these outer boundaries. Table~\ref{tab:strokeStats} lists the test doodles the version of \toolName{} used with participants classified wrongly (W) both after the last stroke and after each stroke. Retraining the doodle classifier after adding these samples yielded better recognition performance (W*). The new classifier performed worse on Camera and Search (and to a lesser degree on Back, Dropdown, Forward, Menu, Play, Plus, Share, and Switch).

Overall recognition accuracy improved from 94.5\% to 94.9\% (while keeping recognition speed the same), but is most notably 9\% better for avatar (the class that study participants had the most trouble with). Among the 186 UI elements in the participants' 34 final screen sketches, the retrained network detected 18 UI elements with fewer strokes (while requiring more strokes for 8 UI elements).

\section{Related Work}

In sketch-based image retrieval (SBIR), computer vision techniques try to find the similarity in the sketch-image pair based on their features when a  user draws an unpolished representation of the image.  Earlier studies extract hand-engineered features (edge-map, Histogram of Oriented Gradients, Histogram of edge local orientation)~\cite{chalechale2004sketch, hu2013performance} to find the similarity between the pair. Deep Leaning achieves state-of-the-art performance in several computer vision applications with Convolutional Neural Network~\cite{AlexNet, VGGNet}. The success also draws researchers to employ deep neural networks for SBIR~\cite{song2017deep,yelamarthi2018zero}. Deep Neural Network(DNN) uses sketch-image pair for training two different networks(one for sketch and one for image).  During the training phase, DNN encodes the image-sketch duo to low-dimensional feature vectors with a target to reduce the distance for similar pairs and maximize for non-similar pairs. For query, it encodes the sketch and then uses the nearest neighbor technique to query similar examples from the dataset. 

Searching design from visual input (image, sketch) recently gaining attention due to the success of DNN and the creation of large-scale datasets.  SWIRE~\cite{huang2019swire} uses a deep neural network model to retrieve relevant UI examples from input sketches.  VINS~\cite{bunian2021vins} UI image (wireframe, high-fidelity) retrieves UI screenshots from high-fidelity wire-frame design. 

In a follow-up to SWIRE, sketching begins with a coarse-level representation of a real-world object, followed by more fine details. Rather than considering a sketch as a flat structure, they use the hierarchical structure to pair it with a photo. Two nodes of the deep neural network are fused to form the next hierarchy level by interacting and matching features between image and sketch pair. They calculated bounding boxes of the individual connected components of the SWIRE drawings to identify interest regions.  A cross-modal co-attention part of the network attends to matching interest regions in a sketch and image pair. By leveraging the hierarchical traits and mutual attention between the interest region, they achieved state-of-the-art performance in the SWIRE dataset.

Successful integration of sketch in the software development process requires a large-scale UI dataset and utilization of the dataset in the deep learning model.  While some freehand drawings of user interface elements are available~\cite{huang2019swire,adefris_Sketch2Code,sermuga2021uisketch}, these sketches are available as ``static'' pixel-based images of the final sketch. 

\balance 

SWIRE~\cite{huang2019swire} collected 3,802 offline sketches of 2,201~screens from 23~app categories of the Google Play store. While the SWIRE dataset is very valuable, it ``only'' contains an offline snapshot of each final UI drawing. And drawings are not tagged with the UI element present in them.  UISketch~\cite{sermuga2021uisketch} introduced the first large-scale dataset of 17,979 hand-drawn sketches of 21 UI element categories collected from 967 participants. 69.38\% of UISketch are digital sketches. The drawings are now publicly available in raw-pixel format with no stroke information.

\section{Conclusions}

Searching through existing repositories for a specific mobile app screen design is currently either slow or tedious. Such searches are either limited to basic keyword searches (Google Image Search) or require as input a complete query screen image (SWIRE). A promising alternative is interactive partial sketching, which is more structured than keyword search and faster than complete-screen queries. \toolName{} is the first system to allow interactive search of screens via interactive sketching. \toolName{} is built on top of a combination of the Rico repository of some 58k Android app screens, the Google \QD{} dataset of icon-level doodles, and \dataName{}, a curated corpus of some 10k app icon doodles collected from hundreds of individuals (mainly crowd-workers). In our evaluation with third-party software developers, \toolName{} provided similar accuracy as the state of the art from the SWIRE line of work, while cutting the average time required about in half. All of \toolName{}'s source code, processing scripts, training data, and experimental results are available under permissive open-source licenses.

\begin{acks}
Christoph Csallner has a potential research conflict of interest due to a financial interest with Microsoft and The Trade Desk. A management plan has been created to preserve objectivity in research in accordance with UTA policy. This material is based upon work supported by the National Science Foundation (NSF) under Grant No. 1911017.
\end{acks}

\bibliographystyle{ACM-Reference-Format}
\bibliography{main}


\begin{thebibliography}{35}


\ifx \showCODEN    \undefined \def \showCODEN     #1{\unskip}     \fi
\ifx \showDOI      \undefined \def \showDOI       #1{#1}\fi
\ifx \showISBNx    \undefined \def \showISBNx     #1{\unskip}     \fi
\ifx \showISBNxiii \undefined \def \showISBNxiii  #1{\unskip}     \fi
\ifx \showISSN     \undefined \def \showISSN      #1{\unskip}     \fi
\ifx \showLCCN     \undefined \def \showLCCN      #1{\unskip}     \fi
\ifx \shownote     \undefined \def \shownote      #1{#1}          \fi
\ifx \showarticletitle \undefined \def \showarticletitle #1{#1}   \fi
\ifx \showURL      \undefined \def \showURL       {\relax}        \fi
\providecommand\bibfield[2]{#2}
\providecommand\bibinfo[2]{#2}
\providecommand\natexlab[1]{#1}
\providecommand\showeprint[2][]{arXiv:#2}

\bibitem[Adefris(2020)]%
        {adefris_Sketch2Code}
\bibfield{author}{\bibinfo{person}{Biniam Adefris}.}
  \bibinfo{year}{2020}\natexlab{}.
\newblock \bibinfo{title}{Sketch2Code}.
\newblock
\newblock
\urldef\tempurl%
\url{https://www.kaggle.com/biniamad/sketch2code}
\showURL{%
\tempurl}


\bibitem[Andersson et~al\mbox{.}(2018)]%
        {Andersson2018SketchCW}
\bibfield{author}{\bibinfo{person}{Melanie Andersson}, \bibinfo{person}{Arvola
  Maja}, {and} \bibinfo{person}{Sara Hedar}.} \bibinfo{year}{2018}\natexlab{}.
\newblock \emph{\bibinfo{title}{Sketch Classification with Neural Networks: A
  Comparative Study of {CNN} and {RNN} on the {Quick, Draw!} data set}}.
\newblock \bibinfo{thesistype}{Master's\ thesis}. \bibinfo{school}{Uppsala
  University}.
\newblock


\bibitem[Buitinck et~al\mbox{.}(2013)]%
        {sklearn_api}
\bibfield{author}{\bibinfo{person}{Lars Buitinck}, \bibinfo{person}{Gilles
  Louppe}, \bibinfo{person}{Mathieu Blondel}, \bibinfo{person}{Fabian
  Pedregosa}, \bibinfo{person}{Andreas Mueller}, \bibinfo{person}{Olivier
  Grisel}, \bibinfo{person}{Vlad Niculae}, \bibinfo{person}{Peter
  Prettenhofer}, \bibinfo{person}{Alexandre Gramfort}, \bibinfo{person}{Jaques
  Grobler}, \bibinfo{person}{Robert Layton}, \bibinfo{person}{Jake VanderPlas},
  \bibinfo{person}{Arnaud Joly}, \bibinfo{person}{Brian Holt}, {and}
  \bibinfo{person}{Ga{\"{e}}l Varoquaux}.} \bibinfo{year}{2013}\natexlab{}.
\newblock \showarticletitle{{API} design for machine learning software:
  experiences from the scikit-learn project}. In \bibinfo{booktitle}{\emph{ECML
  PKDD Workshop: Languages for Data Mining and Machine Learning}}.
  \bibinfo{pages}{108--122}.
\newblock


\bibitem[Bunian et~al\mbox{.}(2021)]%
        {bunian2021vins}
\bibfield{author}{\bibinfo{person}{Sara Bunian}, \bibinfo{person}{Kai Li},
  \bibinfo{person}{Chaima Jemmali}, \bibinfo{person}{Casper Harteveld},
  \bibinfo{person}{Yun Fu}, {and} \bibinfo{person}{Magy~Seif Seif El-Nasr}.}
  \bibinfo{year}{2021}\natexlab{}.
\newblock \showarticletitle{VINS: Visual Search for Mobile User Interface
  Design}. In \bibinfo{booktitle}{\emph{Proceedings of the 2021 CHI Conference
  on Human Factors in Computing Systems}}. \bibinfo{pages}{1--14}.
\newblock


\bibitem[Campos and Nunes(2007)]%
        {Campos2007Practitioner}
\bibfield{author}{\bibinfo{person}{Pedro Campos} {and}
  \bibinfo{person}{Nuno~Jardim Nunes}.} \bibinfo{year}{2007}\natexlab{}.
\newblock \showarticletitle{Practitioner tools and workstyles for
  user-interface design}.
\newblock \bibinfo{journal}{\emph{IEEE software}} \bibinfo{volume}{24},
  \bibinfo{number}{1} (\bibinfo{date}{Jan.} \bibinfo{year}{2007}),
  \bibinfo{pages}{73--80}.
\newblock


\bibitem[Carter and Hundhausen(2010)]%
        {carter2010user}
\bibfield{author}{\bibinfo{person}{Adam~S. Carter} {and}
  \bibinfo{person}{Christopher~D. Hundhausen}.}
  \bibinfo{year}{2010}\natexlab{}.
\newblock \showarticletitle{How is user interface prototyping really done in
  practice? A survey of user interface designers}. In
  \bibinfo{booktitle}{\emph{Proc. {IEEE} Symposium on Visual Languages and
  Human-Centric Computing (VL/HCC)}}. \bibinfo{publisher}{IEEE},
  \bibinfo{pages}{207--211}.
\newblock


\bibitem[Chalechale et~al\mbox{.}(2004)]%
        {chalechale2004sketch}
\bibfield{author}{\bibinfo{person}{Abdolah Chalechale},
  \bibinfo{person}{Golshah Naghdy}, {and} \bibinfo{person}{Alfred Mertins}.}
  \bibinfo{year}{2004}\natexlab{}.
\newblock \showarticletitle{Sketch-based image matching using angular
  partitioning}.
\newblock \bibinfo{journal}{\emph{IEEE Transactions on Systems, Man, and
  Cybernetics-part a: systems and humans}} \bibinfo{volume}{35},
  \bibinfo{number}{1} (\bibinfo{year}{2004}), \bibinfo{pages}{28--41}.
\newblock


\bibitem[Deka et~al\mbox{.}(2017)]%
        {deka2017rico}
\bibfield{author}{\bibinfo{person}{Biplab Deka}, \bibinfo{person}{Zifeng
  Huang}, \bibinfo{person}{Chad Franzen}, \bibinfo{person}{Joshua Hibschman},
  \bibinfo{person}{Daniel Afergan}, \bibinfo{person}{Yang Li},
  \bibinfo{person}{Jeffrey Nichols}, {and} \bibinfo{person}{Ranjitha Kumar}.}
  \bibinfo{year}{2017}\natexlab{}.
\newblock \showarticletitle{Rico: {A} mobile app dataset for building
  data-driven design applications}. In \bibinfo{booktitle}{\emph{Proc. 30th
  Annual {ACM} Symposium on User Interface Software and Technology (UIST)}}.
  \bibinfo{publisher}{ACM}, \bibinfo{pages}{845--854}.
\newblock


\bibitem[Deka et~al\mbox{.}(2016)]%
        {Erica}
\bibfield{author}{\bibinfo{person}{Biplab Deka}, \bibinfo{person}{Zifeng
  Huang}, {and} \bibinfo{person}{Ranjitha Kumar}.}
  \bibinfo{year}{2016}\natexlab{}.
\newblock \showarticletitle{{ERICA:} Interaction mining mobile apps}. In
  \bibinfo{booktitle}{\emph{Proc. 29th Annual Symposium on User Interface
  Software and Technology (UIST)}}. \bibinfo{publisher}{ACM},
  \bibinfo{pages}{767--776}.
\newblock


\bibitem[Eckert and Stacey(2000)]%
        {eckert2000sourceinspiration}
\bibfield{author}{\bibinfo{person}{Claudia Eckert} {and}
  \bibinfo{person}{Martin Stacey}.} \bibinfo{year}{2000}\natexlab{}.
\newblock \showarticletitle{Sources of inspiration: a language of design}.
\newblock \bibinfo{journal}{\emph{Design studies}} \bibinfo{volume}{21},
  \bibinfo{number}{5} (\bibinfo{year}{2000}), \bibinfo{pages}{523--538}.
\newblock


\bibitem[Garrido-Jurado et~al\mbox{.}(2014)]%
        {garrido2014aruco}
\bibfield{author}{\bibinfo{person}{Sergio Garrido-Jurado},
  \bibinfo{person}{Rafael Mu{\~n}oz-Salinas},
  \bibinfo{person}{Francisco~Jos{\'e} Madrid-Cuevas}, {and}
  \bibinfo{person}{Manuel~Jes{\'u}s Mar{\'\i}n-Jim{\'e}nez}.}
  \bibinfo{year}{2014}\natexlab{}.
\newblock \showarticletitle{Automatic generation and detection of highly
  reliable fiducial markers under occlusion}.
\newblock \bibinfo{journal}{\emph{Pattern Recognition}} \bibinfo{volume}{47},
  \bibinfo{number}{6} (\bibinfo{year}{2014}), \bibinfo{pages}{2280--2292}.
\newblock


\bibitem[Ha and Eck(2017)]%
        {ha2017neural}
\bibfield{author}{\bibinfo{person}{David Ha} {and} \bibinfo{person}{Douglas
  Eck}.} \bibinfo{year}{2017}\natexlab{}.
\newblock \showarticletitle{A neural representation of sketch drawings}.
\newblock \bibinfo{journal}{\emph{arXiv preprint arXiv:1704.03477}}.
\newblock


\bibitem[Hellmann and Maurer(2011)]%
        {hellmann2011ruletestUI}
\bibfield{author}{\bibinfo{person}{Theodore~D Hellmann} {and}
  \bibinfo{person}{Frank Maurer}.} \bibinfo{year}{2011}\natexlab{}.
\newblock \showarticletitle{Rule-based exploratory testing of graphical user
  interfaces}. In \bibinfo{booktitle}{\emph{2011 Agile Conference}}. IEEE,
  \bibinfo{pages}{107--116}.
\newblock


\bibitem[Herring et~al\mbox{.}(2009)]%
        {herring2009designpractise}
\bibfield{author}{\bibinfo{person}{Scarlett~R Herring},
  \bibinfo{person}{Chia-Chen Chang}, \bibinfo{person}{Jesse Krantzler}, {and}
  \bibinfo{person}{Brian~P Bailey}.} \bibinfo{year}{2009}\natexlab{}.
\newblock \showarticletitle{Getting inspired! Understanding how and why
  examples are used in creative design practice}. In
  \bibinfo{booktitle}{\emph{Proceedings of the SIGCHI conference on human
  factors in computing systems}}. \bibinfo{pages}{87--96}.
\newblock


\bibitem[Hu and Collomosse(2013)]%
        {hu2013performance}
\bibfield{author}{\bibinfo{person}{Rui Hu} {and} \bibinfo{person}{John
  Collomosse}.} \bibinfo{year}{2013}\natexlab{}.
\newblock \showarticletitle{A performance evaluation of gradient field hog
  descriptor for sketch based image retrieval}.
\newblock \bibinfo{journal}{\emph{Computer Vision and Image Understanding}}
  \bibinfo{volume}{117}, \bibinfo{number}{7} (\bibinfo{year}{2013}),
  \bibinfo{pages}{790--806}.
\newblock


\bibitem[Huang et~al\mbox{.}(2019)]%
        {huang2019swire}
\bibfield{author}{\bibinfo{person}{Forrest Huang}, \bibinfo{person}{John~F.
  Canny}, {and} \bibinfo{person}{Jeffrey Nichols}.}
  \bibinfo{year}{2019}\natexlab{}.
\newblock \showarticletitle{Swire: Sketch-based user interface retrieval}. In
  \bibinfo{booktitle}{\emph{Proceedings of the 2019 CHI Conference on Human
  Factors in Computing Systems}}. \bibinfo{publisher}{ACM}.
\newblock


\bibitem[Huffman and Hochster(2007)]%
        {huffman2007well}
\bibfield{author}{\bibinfo{person}{Scott~B Huffman} {and}
  \bibinfo{person}{Michael Hochster}.} \bibinfo{year}{2007}\natexlab{}.
\newblock \showarticletitle{How well does result relevance predict session
  satisfaction?}. In \bibinfo{booktitle}{\emph{Proc. 30th Annual International
  {ACM} {SIGIR} Conference on Research and Development in Information
  Retrieval}}. \bibinfo{publisher}{ACM}, \bibinfo{pages}{567--574}.
\newblock


\bibitem[Ines et~al\mbox{.}(2017)]%
        {ines2017evalmobileinterface}
\bibfield{author}{\bibinfo{person}{Gasmi Ines}, \bibinfo{person}{Soui Makram},
  \bibinfo{person}{Chouchane Mabrouka}, {and} \bibinfo{person}{Abed Mourad}.}
  \bibinfo{year}{2017}\natexlab{}.
\newblock \showarticletitle{Evaluation of mobile interfaces as an optimization
  problem}.
\newblock \bibinfo{journal}{\emph{Procedia computer science}}
  \bibinfo{volume}{112} (\bibinfo{year}{2017}), \bibinfo{pages}{235--248}.
\newblock


\bibitem[Jongejan et~al\mbox{.}(2016)]%
        {jongejan2016quick}
\bibfield{author}{\bibinfo{person}{Jonas Jongejan}, \bibinfo{person}{Henry
  Rowley}, \bibinfo{person}{Takashi Kawashima}, \bibinfo{person}{Jongmin Kim},
  {and} \bibinfo{person}{Nick Fox-Gieg}.} \bibinfo{year}{2016}\natexlab{}.
\newblock \bibinfo{title}{Quick, Draw!}
\newblock
\newblock
\urldef\tempurl%
\url{https://quickdraw.withgoogle.com/}
\showURL{%
\tempurl}
\newblock
\shownote{Accessed March 2022}.


\bibitem[Krizhevsky et~al\mbox{.}(2012)]%
        {AlexNet}
\bibfield{author}{\bibinfo{person}{Alex Krizhevsky}, \bibinfo{person}{Ilya
  Sutskever}, {and} \bibinfo{person}{Geoffrey~E. Hinton}.}
  \bibinfo{year}{2012}\natexlab{}.
\newblock \showarticletitle{{ImageNet} classification with deep convolutional
  neural networks}. In \bibinfo{booktitle}{\emph{Proc. 26th Annual Conference
  on Neural Information Processing Systems (NIPS)}}. \bibinfo{publisher}{NIPS},
  \bibinfo{pages}{1106--1114}.
\newblock


\bibitem[Landay and Myers(1995)]%
        {Landay95Interactive}
\bibfield{author}{\bibinfo{person}{James~A. Landay} {and}
  \bibinfo{person}{Brad~A. Myers}.} \bibinfo{year}{1995}\natexlab{}.
\newblock \showarticletitle{Interactive sketching for the early stages of user
  interface design}. In \bibinfo{booktitle}{\emph{Proc. ACM SIGCHI Conference
  on Human Factors in Computing Systems (CHI)}}. \bibinfo{publisher}{ACM},
  \bibinfo{pages}{43--50}.
\newblock


\bibitem[Liu et~al\mbox{.}(2018)]%
        {liu2018learning}
\bibfield{author}{\bibinfo{person}{Thomas~F Liu}, \bibinfo{person}{Mark Craft},
  \bibinfo{person}{Jason Situ}, \bibinfo{person}{Ersin Yumer},
  \bibinfo{person}{Radomir Mech}, {and} \bibinfo{person}{Ranjitha Kumar}.}
  \bibinfo{year}{2018}\natexlab{}.
\newblock \showarticletitle{Learning design semantics for mobile apps}. In
  \bibinfo{booktitle}{\emph{Proc. 31st Annual ACM Symposium on User Interface
  Software and Technology (UIST)}}. \bibinfo{pages}{569--579}.
\newblock


\bibitem[Mohian and Csallner(2021)]%
        {soumik_mohian_DoodleUINet}
\bibfield{author}{\bibinfo{person}{Soumik Mohian} {and}
  \bibinfo{person}{Christoph Csallner}.} \bibinfo{year}{2021}\natexlab{}.
\newblock \bibinfo{booktitle}{\emph{{DoodleUINet: Repository for DoodleUINet
  Drawings Dataset and Scripts}}}.
\newblock
\urldef\tempurl%
\url{https://doi.org/10.5281/zenodo.5144472}
\showDOI{\tempurl}


\bibitem[Mohian and Csallner(2022)]%
        {soumikmohianuta_psdoodle_repo}
\bibfield{author}{\bibinfo{person}{Soumik Mohian} {and}
  \bibinfo{person}{Christoph Csallner}.} \bibinfo{year}{2022}\natexlab{}.
\newblock \bibinfo{booktitle}{\emph{{soumikmohianuta/PSDoodle: PSDoodle
  Repository for the Publication}}}.
\newblock
\urldef\tempurl%
\url{https://doi.org/10.5281/zenodo.6339717}
\showDOI{\tempurl}


\bibitem[Newman and Landay(1999)]%
        {Newman99Sitemaps}
\bibfield{author}{\bibinfo{person}{Mark~W. Newman} {and}
  \bibinfo{person}{James~A. Landay}.} \bibinfo{year}{1999}\natexlab{}.
\newblock \bibinfo{booktitle}{\emph{Sitemaps, storyboards, and specifications:
  A sketch of {Web} site design practice as manifested through artifacts}}.
\newblock \bibinfo{type}{{T}echnical {R}eport} UCB/CSD-99-1062.
  \bibinfo{institution}{EECS Department, UC Berkeley}.
\newblock


\bibitem[Ritchie et~al\mbox{.}(2011)]%
        {ritchie2011d}
\bibfield{author}{\bibinfo{person}{Daniel Ritchie},
  \bibinfo{person}{Ankita~Arvind Kejriwal}, {and} \bibinfo{person}{Scott~R
  Klemmer}.} \bibinfo{year}{2011}\natexlab{}.
\newblock \showarticletitle{d. tour: Style-based exploration of design example
  galleries}. In \bibinfo{booktitle}{\emph{Proc. 24th annual ACM Symposium on
  User Interface Software and Technology (UIST)}}. \bibinfo{pages}{165--174}.
\newblock


\bibitem[Sain et~al\mbox{.}(2020)]%
        {sain2020cross}
\bibfield{author}{\bibinfo{person}{Aneeshan Sain}, \bibinfo{person}{Ayan~Kumar
  Bhunia}, \bibinfo{person}{Yongxin Yang}, \bibinfo{person}{Tao Xiang}, {and}
  \bibinfo{person}{Yi-Zhe Song}.} \bibinfo{year}{2020}\natexlab{}.
\newblock \showarticletitle{Cross-modal hierarchical modelling for fine-grained
  sketch based image retrieval}. In \bibinfo{booktitle}{\emph{Proc. 31st
  British Machine Vision Virtual Conference (BMVC)}}.
\newblock


\bibitem[Sermuga~Pandian et~al\mbox{.}(2021)]%
        {sermuga2021uisketch}
\bibfield{author}{\bibinfo{person}{Vinoth~Pandian Sermuga~Pandian},
  \bibinfo{person}{Sarah Suleri}, {and} \bibinfo{person}{Prof Dr~Matthias
  Jarke}.} \bibinfo{year}{2021}\natexlab{}.
\newblock \showarticletitle{UISketch: A Large-Scale Dataset of UI Element
  Sketches}. In \bibinfo{booktitle}{\emph{Proceedings of the 2021 CHI
  Conference on Human Factors in Computing Systems}}. \bibinfo{pages}{1--14}.
\newblock


\bibitem[Simonyan and Zisserman(2015)]%
        {VGGNet}
\bibfield{author}{\bibinfo{person}{Karen Simonyan} {and}
  \bibinfo{person}{Andrew Zisserman}.} \bibinfo{year}{2015}\natexlab{}.
\newblock \showarticletitle{Very Deep Convolutional Networks for Large-Scale
  Image Recognition}. In \bibinfo{booktitle}{\emph{arXiv:1409.1556}}.
\newblock


\bibitem[Song et~al\mbox{.}(2017)]%
        {song2017deep}
\bibfield{author}{\bibinfo{person}{Jifei Song}, \bibinfo{person}{Qian Yu},
  \bibinfo{person}{Yi-Zhe Song}, \bibinfo{person}{Tao Xiang}, {and}
  \bibinfo{person}{Timothy~M Hospedales}.} \bibinfo{year}{2017}\natexlab{}.
\newblock \showarticletitle{Deep spatial-semantic attention for fine-grained
  sketch-based image retrieval}. In \bibinfo{booktitle}{\emph{Proc. IEEE
  International Conference on Computer Vision}}. \bibinfo{pages}{5551--5560}.
\newblock


\bibitem[Tensorflow(2020)]%
        {quickdraw_rnn}
\bibfield{author}{\bibinfo{person}{Tensorflow}.}
  \bibinfo{year}{2020}\natexlab{}.
\newblock \bibinfo{title}{Recurrent Neural Networks for Drawing
  Classification}.
\newblock
\newblock
\urldef\tempurl%
\url{https://github.com/tensorflow/docs/blob/master/site/en/r1/tutorials/sequences/recurrent_quickdraw.md}
\showURL{%
\tempurl}


\bibitem[Torrey and Shavlik(2009)]%
        {torrey2010transfer}
\bibfield{author}{\bibinfo{person}{Lisa Torrey} {and} \bibinfo{person}{Jude
  Shavlik}.} \bibinfo{year}{2009}\natexlab{}.
\newblock \showarticletitle{Transfer learning}.
\newblock In \bibinfo{booktitle}{\emph{Handbook of Research on Machine Learning
  Applications and Trends: Algorithms, Methods, and Techniques}}.
  \bibinfo{publisher}{IGI Global}, \bibinfo{pages}{242--264}.
\newblock


\bibitem[Wong(1992)]%
        {Wong92Rough}
\bibfield{author}{\bibinfo{person}{Yin~Yin Wong}.}
  \bibinfo{year}{1992}\natexlab{}.
\newblock \showarticletitle{Rough and ready prototypes: Lessons from graphic
  design}. In \bibinfo{booktitle}{\emph{Proc. ACM SIGCHI Conference on Human
  Factors in Computing Systems (CHI), Posters and Short Talks}}.
  \bibinfo{publisher}{ACM}, \bibinfo{pages}{83--84}.
\newblock


\bibitem[Yeh et~al\mbox{.}(2009)]%
        {yeh2009sikuli}
\bibfield{author}{\bibinfo{person}{Tom Yeh}, \bibinfo{person}{Tsung-Hsiang
  Chang}, {and} \bibinfo{person}{Robert~C Miller}.}
  \bibinfo{year}{2009}\natexlab{}.
\newblock \showarticletitle{Sikuli: using GUI screenshots for search and
  automation}. In \bibinfo{booktitle}{\emph{Proceedings of the 22nd annual ACM
  symposium on User interface software and technology}}.
  \bibinfo{pages}{183--192}.
\newblock


\bibitem[Yelamarthi et~al\mbox{.}(2018)]%
        {yelamarthi2018zero}
\bibfield{author}{\bibinfo{person}{Sasi~Kiran Yelamarthi},
  \bibinfo{person}{Shiva~Krishna Reddy}, \bibinfo{person}{Ashish Mishra}, {and}
  \bibinfo{person}{Anurag Mittal}.} \bibinfo{year}{2018}\natexlab{}.
\newblock \showarticletitle{A zero-shot framework for sketch based image
  retrieval}. In \bibinfo{booktitle}{\emph{Proceedings of the European
  Conference on Computer Vision (ECCV)}}. \bibinfo{pages}{300--317}.
\newblock


\end{thebibliography}

\appendix

\end{document}